\documentclass[3p,times,procedia]{elsarticle}
\usepackage{amssymb}
\usepackage{lineno}
\usepackage[figuresright]{rotating}

\newcommand{\pt}{\ensuremath{p_{\mathrm{T}}}}
\newcommand{\ph}{\ensuremath{\phi}}
\newcommand{\pion}{\ensuremath{\pi}}

\newcommand{\snn}{\ensuremath{\sqrt{s_{\mathrm{NN}}}}}

\newcommand{\dndeta}{\ensuremath{\langle \mathrm{d}N_{\mathrm{ch}}/\mathrm{d}\eta\rangle}}
\newcommand{\meanpt}{\ensuremath{\langle \pt \rangle}}

\begin{document}

\begin{frontmatter}
\begin{center}
XXVIIth International Conference on Ultrarelativistic Nucleus-Nucleus Collisions (Quark Matter 2018)
\end{center}

\title{Testing the system size dependence of hydrodynamical expansion and thermal particle production with $\pi$, K, p, and $\phi$ in Xe--Xe and Pb--Pb collisions with ALICE}
\author[label1]{Francesca Bellini (for the ALICE Collaboration)} 
\address[label1]{European Organization for Nuclear Research (CERN), Geneva, Switzerland \\ E-mail: francesca.bellini@cern.ch}

\begin{abstract}
We present new results on transverse momentum spectra, integrated yields, and mean transverse momenta of pions, kaons, and protons, as well as of $\phi$-mesons for various centrality classes measured in Pb--Pb and Xe--Xe collisions at the LHC. This unique set of data allows us to investigate bulk particle production for very different systems at similar multiplicities. The chemical and kinetic freeze-out parameters are extracted via statistical-thermal and combined blast-wave fits to the data in heavy-ion collisions and are compared to results obtained in pp and p--Pb collisions at similar multiplicities. The evolution of collective-like effects from pp and p--Pb collisions to Xe--Xe and Pb--Pb collisions is further investigated by detailed comparisons to predictions from models. 
\end{abstract}
 
\begin{keyword}
Identified hadrons yields \sep particle ratios \sep $\phi$-meson \sep thermal fit \sep hydrodynamics \sep Xe--Xe \sep Pb--Pb
\end{keyword}

\end{frontmatter}


\section{Introduction}
At the LHC, ALICE has performed a comprehensive set of measurements of light-flavour hadron production in different collision systems and at various energies, unveiling (a.) a continuous evolution of relative particle yields across collision systems that seems to depend only on charged particle multiplicity regardless of collision energy and system type \cite{ALICE:2017jyt}, and (b.) the presence of collective-like effects in the way the maxima of the transverse momentum ($p_{\rm{T}}$) dependent spectra and baryon-to-meson ratios exhibit a blueshift going from low to high multiplicity events, even in small systems \cite{Acharya:2018orn}. In heavy-ion collisions, measurements of the relative abundances of light-flavour hadrons can be used to infer the properties of the system at chemical freeze-out, such as the temperature, $T_{chem}$. Particle \pt~spectra, mean \pt~and \pt-differential baryon-to-meson ratios, determined at the kinetic freeze-out, can be used to test hydrodynamic models and the system-scaling properties of their assumptions \cite{Giacalone:2017dud}. 

Preliminary results on the production of \pion, K, p, and $\phi$ in Pb--Pb collisions at $\snn = 5.02$ TeV were reported in \cite{Jacazio:2018wdf, Fragiacomo:2018sdk}. 
During a six-hours LHC pilot run in November 2017, ALICE recorded Xe--Xe collisions at $\snn = 5.44$ TeV. A sample of about $1.4 \times 10^{6}$ minimum-bias triggered events was analysed to measure the production of \pion, K, p and $\phi$-meson at mid-rapidity, following a similar strategy as for the 5.02 TeV Pb--Pb data. 
Primary charged \pion, K and p are tracked and identified in the ALICE central barrel using the Inner Tracking System (ITS), the Time Projection Chamber (TPC) and the Time-Of-Flight (TOF) detector. 
The $\phi$-meson is reconstructed with an invariant mass analysis via the decay channel $\phi\rightarrow \mathrm{K}^{+}\mathrm{K}^{-}$, where charged kaons are identified with TPC and TOF. 
Centrality classes are defined based on the signal amplitude in the V0 scintillators placed at forward rapidity, whereas for each class, the average charged particle multiplicity density, \dndeta, is measured in $\vert \eta \vert < 0.5$ \cite{Acharya:2018hhy}.
In Xe--Xe collisions, \pion, K and p are measured in nine centrality intervals and in the following \pt~ranges: 0.15 - 5 GeV/$c$ for \pion, 0.2 - 3.6 GeV/$c$ for K, 0.3 - 5 GeV/$c$ for p. The $\phi$-meson is measured in four centrality classes, for $0.3 < \pt < 10$ GeV/$c$. 

\section{Testing hydrodynamics with identified hadron data}
\begin{figure}[ht]
\begin{center}
\includegraphics[width=.365\textwidth]{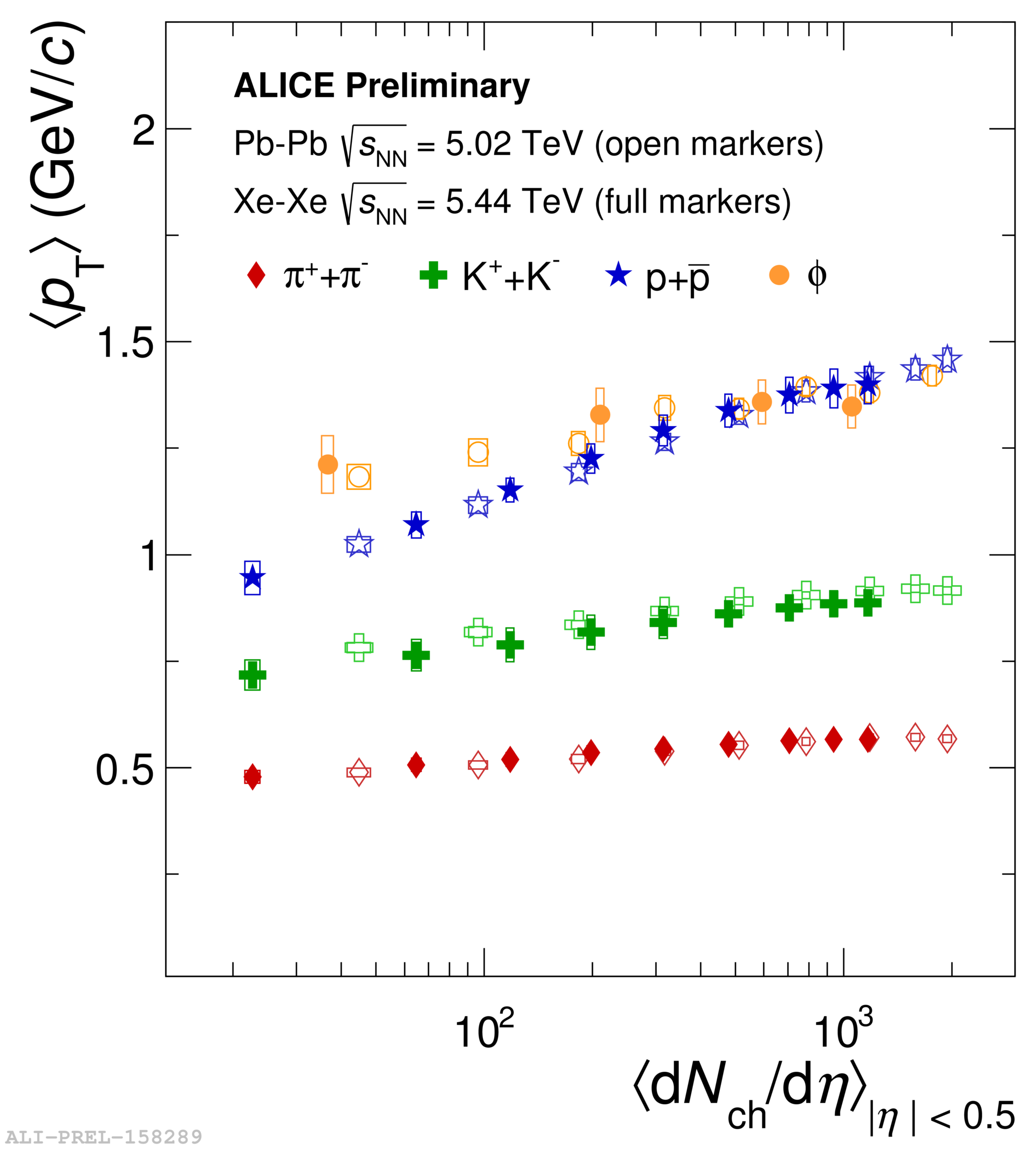}
\includegraphics[width=.43\textwidth]{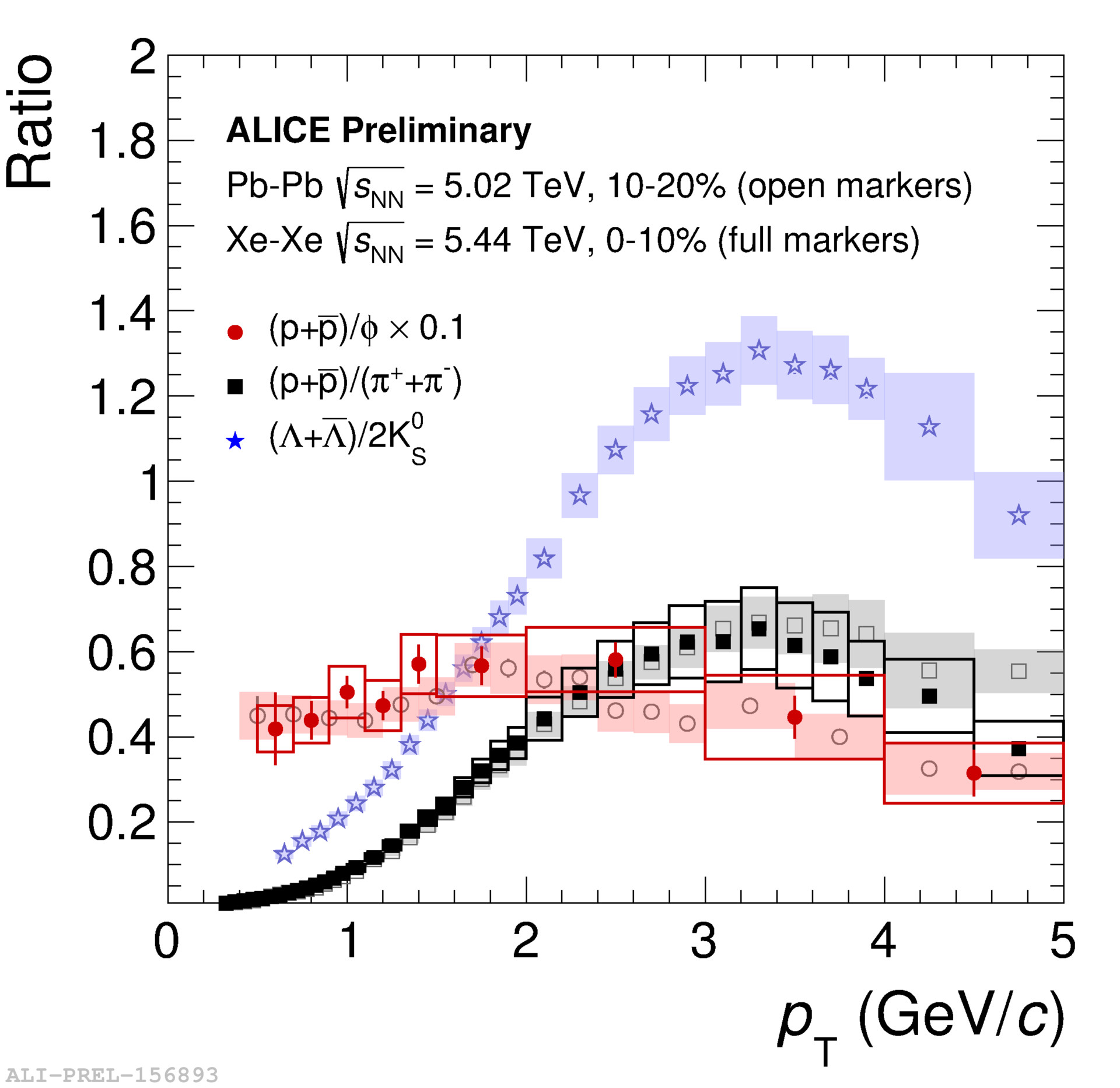}
\caption{Left: Multiplicity dependence of the average transverse momentum for identified hadrons in Pb--Pb and Xe--Xe collisions. Right: \pt-dependent baryon-to-meson ratios in Xe--Xe (0-10$\%$) and Pb--Pb (10-20$\%$) collisions compared for centrality classes that have similar charged particle multiplicity density. In both panels, statistical and systematic uncertainties are represented as bars and boxes, respectively.}
\label{fig:meanpt}
\end{center} 
\end{figure}

In Pb--Pb and Xe--Xe collisions, the transverse momentum spectra of the measured hadrons become harder with increasing centrality. This is mirrored in the increasing trend of the average transverse momentum (\meanpt) with centrality, reported in Fig. \ref{fig:meanpt} (left). The \meanpt~of identified hadrons follows mass ordering in central collisions, being larger for particles with larger mass and similar for particles with similar mass, such as the p and the \ph. These observations are consistent with expectations from hydrodynamics, as particles in the expanding system experience the same radial velocity field. The new data in Xe--Xe and in Pb--Pb collisions provide a consistent picture and evidence a scaling of \meanpt~with the charged particle multiplicity, as also observed for inclusive charged hadrons \cite{Acharya:2018eaq}. Hydrodynamic calculations \cite{Giacalone:2017dud} predict a difference in \meanpt~of the order of 2$\%$ between Xe--Xe and Pb--Pb, which is consistent with the observations, given the present uncertainties. 
\noindent In addition, the measured \pt-differential p/\pion~(Fig. \ref{fig:meanpt}, right), K/\pion~(not shown here) and p/\ph~ratios (Fig. \ref{fig:meanpt}, right) in 5.02 TeV Pb--Pb collisions and 5.44 TeV Xe--Xe are consistent within uncertainties once compared at the same multiplicity \dndeta. 
Baryon-to-meson ratios constitute an important input to study particle production mechanisms at intermediate \pt, where both radial flow and recombination play a role in determining the particle spectra. 
On one hand, the enhancement of proton and $\Lambda$ over \pion~and K$^{0}_{\mathrm{S}}$ respectively, is understood as due to radial flow. 
On the other hand, the flatness of the p/\ph~ratio is consistent with hydrodynamics expectations (particles with similar mass have similar spectral shapes) and, at the same time, it is reproduced by models with recombination \cite{Minissale:2015zwa}. 

In order to characterise the kinetic freeze-out and try to quantify radial flow in different systems, the Boltzmann-Gibbs blast-wave model \cite{Schnedermann:1993ws} is commonly employed to fit simultaneously the \pt-spectra of \pion, K and p and extract the kinetic freeze-out temperature, $T_{kin}$ and the radial expansion velocity, $\beta_{\mathrm{T}}$ of the system \cite{Abelev:2013vea}. In Pb--Pb and Xe--Xe collisions, $\beta_{\mathrm{T}}$ increases while $T_{kin}$ decreases with increasing centrality. The fits to Pb--Pb and Xe--Xe data result in parameters that are consistent at similar \dndeta, once again highlighting how their evolution seems to depend only on multiplicity, regardless of the type of the colliding nucleus.  
At similar multiplicities, $\beta_{\mathrm{T}}$ is larger for small systems than for heavy-ion collisions, see \cite{Acharya:2018orn}. 
Being a simplified hydrodynamics-inspired model, the blast-wave fit does not substitute the comparison with full hydrodynamics calculations, yet
the fits remain a useful tool to compare radial flow parameters in different systems, as long as one is full aware of the caveats, as discussed in \cite{Bellini:2017onh}. 

Data from Pb--Pb collisions have been compared to different models, as reported in \cite{Jacazio:2018wdf}. It is observed that models based on viscous hydrodynamics with different sets of initial conditions (iEbyE + VISHNU with Trento or AMPT initial conditions \cite{Zhao:2017yhj, Bhalerao:2015iya}, MUSIC with IP-Glasma initial conditions \cite{McDonald:2016vlt}) reproduce features of particle spectra and particle ratios in central Pb--Pb collisions for $\pt < 2$ GeV/$c$ at the level of 20-30$\%$. 
EPOS-LHC \cite{Pierog:2013ria} does not reproduce satisfactorily individual \pion, K and p spectra in central Pb--Pb collisions, although it describes qualitatively the \pt-differential particle ratios. 
For the tested models, the agreement with data worsens towards peripheral events.

In summary, the comparison of the new Xe--Xe data with the preliminary results in Pb--Pb at $\snn = $ 5.02 TeV does not evidence any significant deviation from the expectations from hydrodynamics, which is confirmed as a valid description for AA collisions.  At intermediate \pt, hydrodynamics breaks down.
If recombination or flow or a combination of both determines the spectral shapes at intermediate \pt~remains an open point, whose investigation could benefit from an increased precision in data and from the direct comparison with model calculations.

\section{Identified hadron yields and thermal model fit to the 5.02 TeV Pb--Pb data} 

\begin{figure}[htbp]
\begin{center}
\includegraphics[width=.52\textwidth]{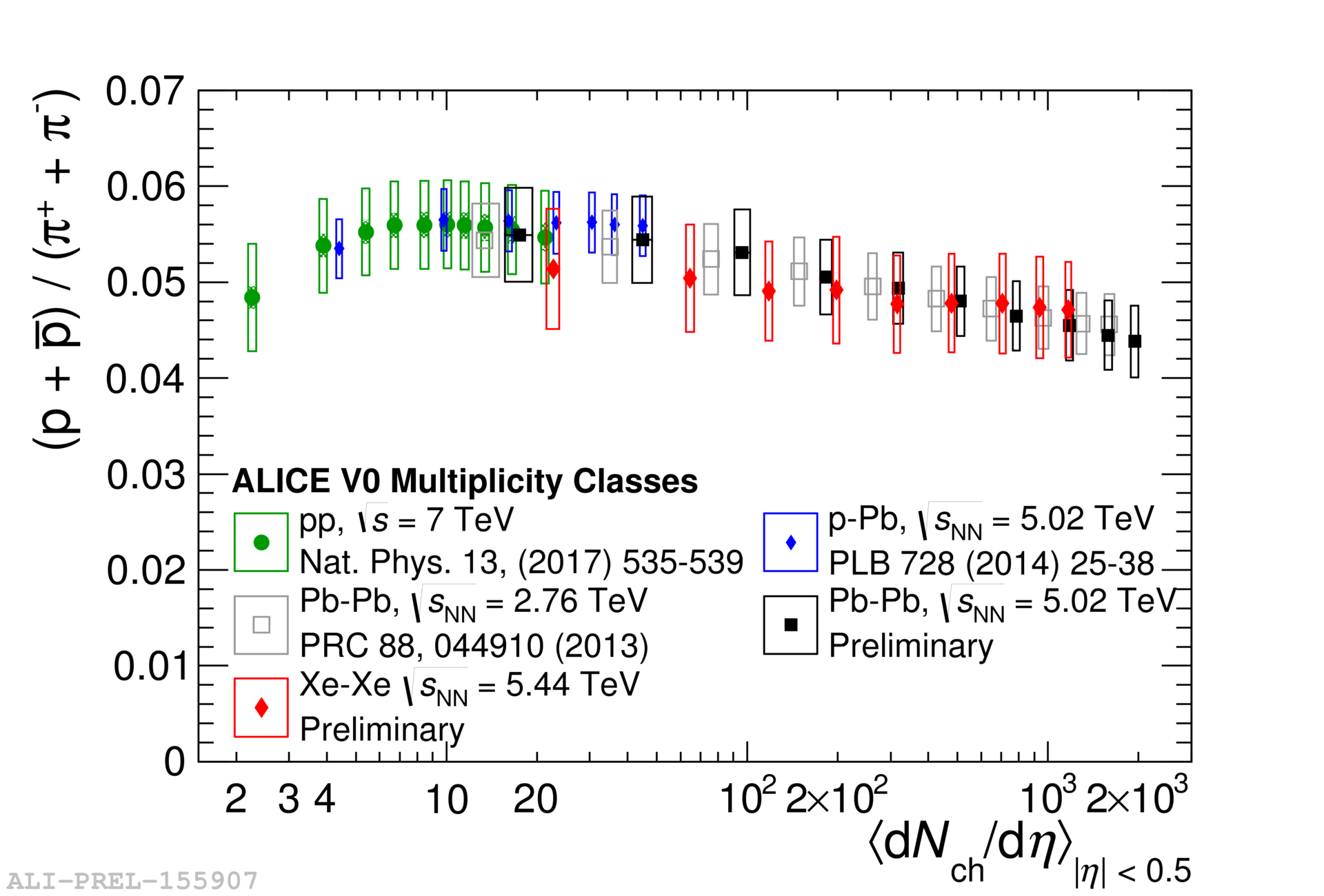}
\includegraphics[width=.47\textwidth]{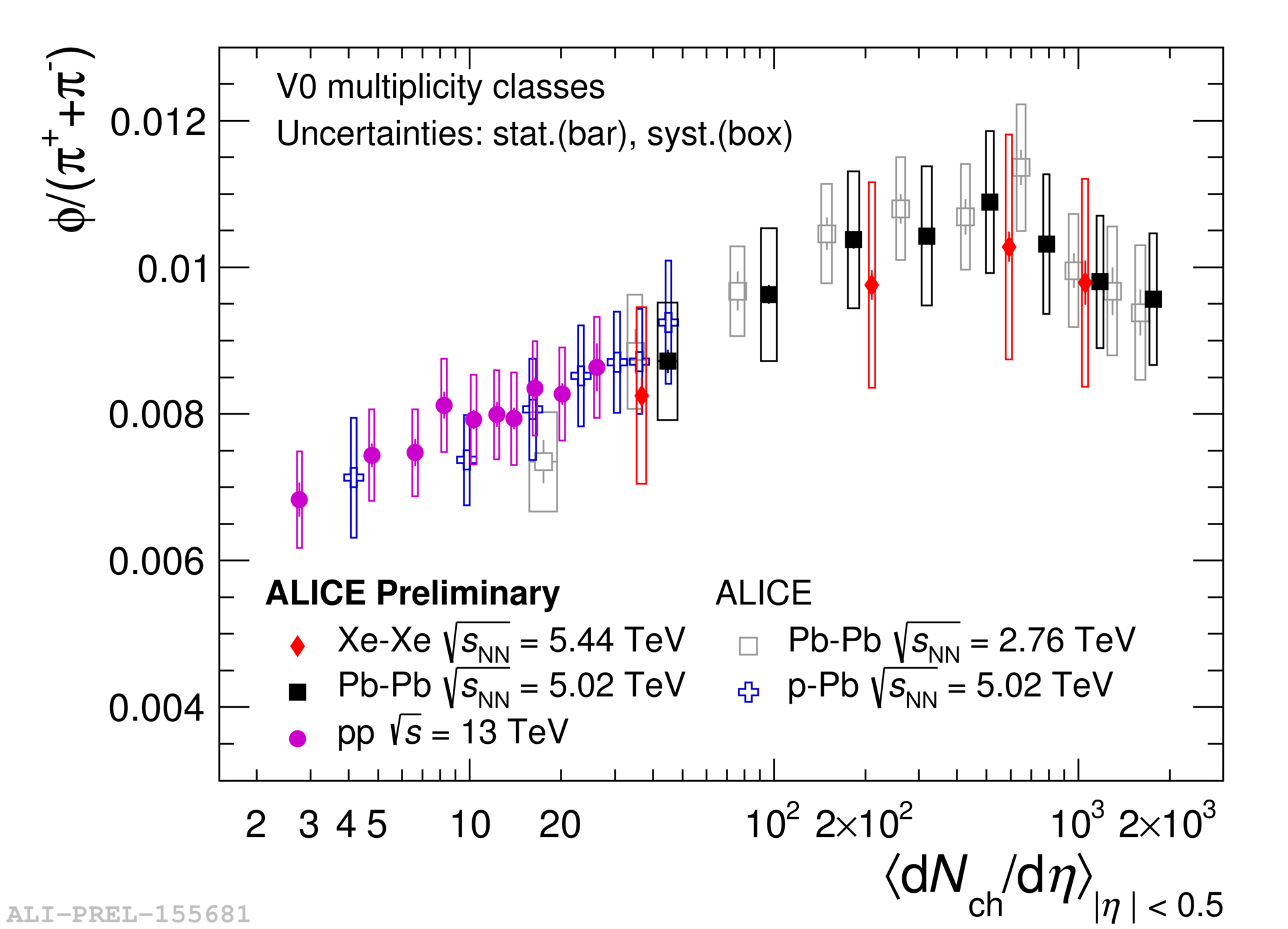}
\caption{\pt-integrated p/\pion~(left) and \ph/\pion~(right) ratios as a function of \dndeta~in different collision systems measured by ALICE. In both panels, statistical and systematic uncertainties are represented as bars and boxes, respectively.}
\label{fig:ratios}
\end{center}
\end{figure}

The \pt-integrated p/\pion~and \ph/\pion~ratios are reported in Fig. \ref{fig:ratios} as a function of \dndeta~in different collision systems. Ratios in AA collisions are consistent at similar multiplicity, independent of collision system (Xe--Xe or Pb--Pb) or energy (\snn~= 5.44 TeV, 5.02 TeV or 2.76 TeV). A similar behaviour is also observed for the K/\pion~ratio. 

\begin{figure}[htbp]
\begin{center}
\includegraphics[width=.8\textwidth]{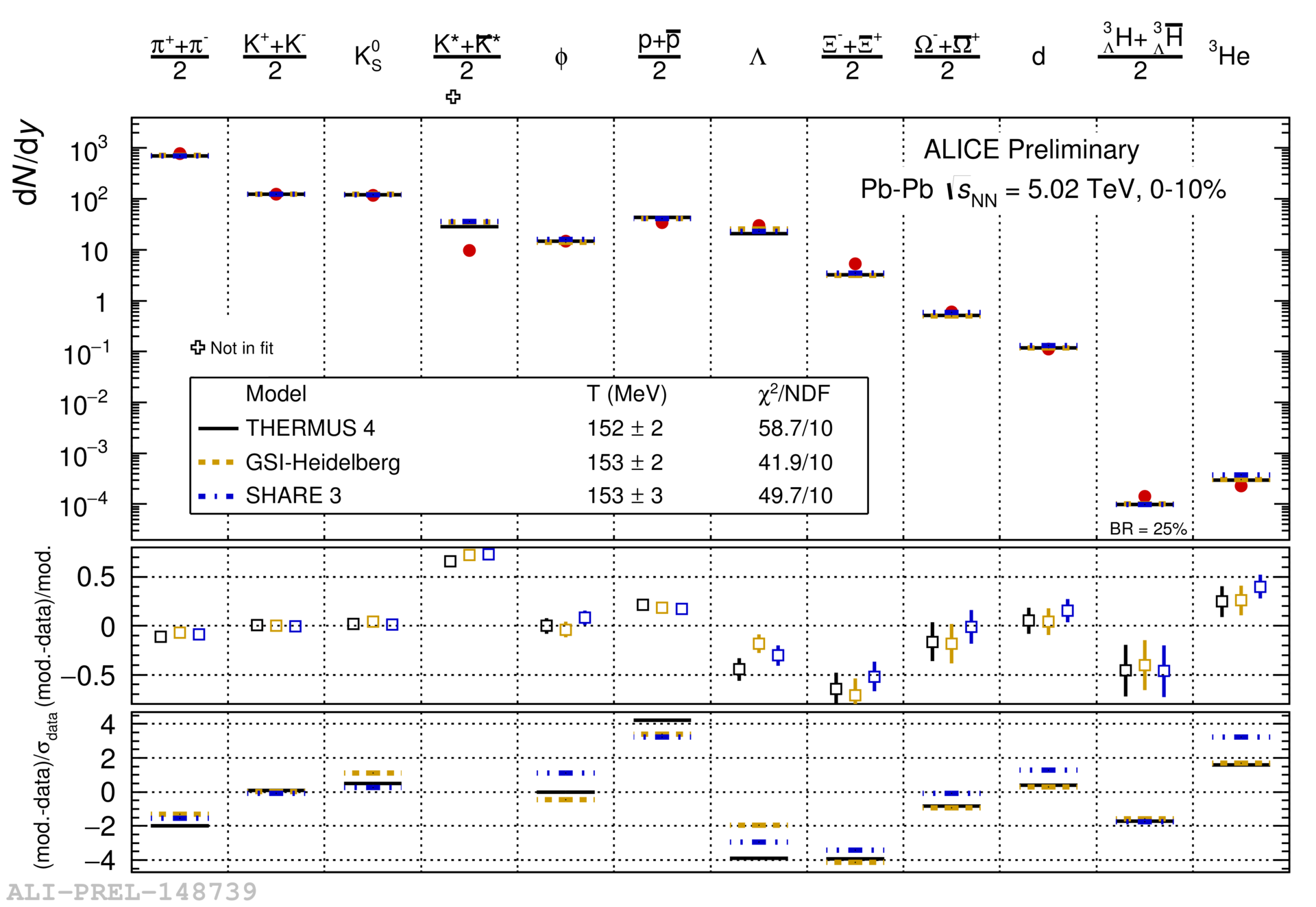}
\caption{Results of the thermal model fit to the yields of identified hadrons measured by ALICE in central (0-10$\%$) Pb--Pb collisions at $\snn =$ 5.02 TeV. Data are preliminary results.}
\label{fig:thermal}
\end{center}
\end{figure}

At the LHC, in Pb--Pb collisions at $\snn = 2.76$ TeV, the production of most light-flavour hadrons and light (anti-)(hyper-)nuclei can be described by thermal models with a single chemical freeze-out temperature, $T_{ch} \approx 156$ MeV \cite{Acharya:2017bso}. 
Thermal model fits to the preliminary ALICE data for the yields of \pion, K, \ph, p, $\Lambda$, $\Xi$, $\Omega$, d, $ ^{3}_{\Lambda}$H and $ ^{3}$He (see Fig. \ref{fig:thermal}) measured in Pb--Pb collisions at $\snn =$ 5.02 TeV converge with $T_{ch} \approx 153$ MeV and $\chi^{2}/N_{dof} \approx 4-6$. 
The three different implementations of the thermal model that we considered provide fully consistent results. The differences between data and model values observed at 2.76 TeV are confirmed at the new energy. 
Deviations from model predictions for the short-lived K$^{*0}$ resonance are ascribed to the presence of re-scattering effects in the hadronic phase. Several hypotheses (an incomplete hadron spectrum considered in the model \cite{Stachel:2013zma}, baryon-antibaryons annihilation in the hadronic phase \cite{Becattini:2014hla, StockQM18}, the effect of finite resonance widths \cite{Vovchenko:2018fmh, Andronic:2018qqt}) have been brought forward to explain the tension between protons and multi-strange hadrons. Potential breakthroughs have been discussed at this conference, which however do not find (yet) unanimous consent in the community, thus requiring further investigation in the near future.

In summary, the newest ALICE results confirm the observation of a smooth evolution of particle composition from pp to p--Pb to AA. The thermal fits to the 5.02 TeV Pb--Pb data lead to a slightly lower chemical freeze-out temperature than at 2.76 TeV. Differences between data and models in the proton and strangeness sector need further investigation and understanding. The precision and the large amount of data available indicate that we are entering an era of precision tests for thermal models. 

\end{document}